# Economic and Energetic Assessment of a Hybrid Vanadium Redox Flow and Lithium-ion batteries, considering Power Sharing Strategies Impact

Ana Foles[a,b,1], Luís Fialho[a,b,2], Pedro Horta[a,b,3], Manuel Collares-Pereira[a,b,4]

[a]*Renewable Energies Chair, University of Évora. Pólo da Mitra da Universidade de Évora, Edifício Ário Lobo de Azevedo, 7000-083 Nossa Senhora da Tourega, Portugal*
[b]*Institute of Earth Sciences, University of Évora, Rua Romão Ramalho, 7000-671, Évora, Portugal*
[1]anafoles@uevora.pt
[2]lafialho@uevora.pt
[3]phorta@uevora.pt
[4]collarespereira@uevora.pt

## Abstract

Hybrid energy storage systems (HESS) combine different energy storage technologies aiming at overall system performance and lifetime improvement compared to a single technology system. In this work, control combinations for a vanadium redox flow battery (VRFB, 5/60 kW/kWh) and a lithium-ion battery (LIB, 3.3/9.8 kW/kWh) are investigated for the design of a HESS. A literature review presents the available energy management/power allocation options that are being studied and applied worldwide in batteries. There is an identified need for opportunities to address better HESS configuration's economic and energy perspective for building applications. The justification of investment in such HESS should improve indicators on use scenarios based on energy management compared to single-battery scenarios. In that context, four scenarios for real-time algorithms application approaches are considered to operate the hybrid storage solution through a 15-year economic and energetic analysis using experimentally validated battery performance models. The results obtained for each scenario are compared with a single technology battery performance to analyse this HESS pair competitiveness and the relevance of the power-sharing techniques among the different ESS technologies, which should be weighted. In the definition of the scenarios, real electricity generation is considered from two solar photovoltaic installations (3.2 kWp and 6.7 kWp) and an estimated representative load of a services building. HESS performance is evaluated through specific energy and economic key performance indicators. The results indicate that using customised energy management strategies (EMSs) renders the VRFB and LIB characteristics complementary, besides enhancing the competitiveness of VRFB as a single technology. Moreover, the HESS management impacts the seasonality factor, contributing to the overall electric system smart management.

| Nomenclature | |
|---|---|
| *Abbreviation, Definition* | |
| BCR | Battery Charge Ratio |
| BESS | Battery Energy Storage System |
| BMS | Battery Management System |
| BTN | Normal Low Voltage |
| CEP | (European) Clean Energy Package |
| DOD | Depth of Discharge (%) |
| DSO | Distributor System Operator |
| EG | Energy from the Grid |
| EMS | Energy Management Strategy |
| ESS | Energy Storage System |
| EV | Electric Vehicle |
| EOL | End of life |
| FBU | From Battery Use |
| FGU | From Grid Use |
| FL | Fuzzy Logic |
| GRF | Grid Relief Factor |
| HESS | Hybrid Energy Storage System |
| IEA | International Energy Agency |
| IRR | Internal Rate of Return (%) |
| KPI | Key-Performance Indicator |
| LCOE | Levelized Cost of Energy (€/kWh) |
| LIB | Lithium-Ion Battery |
| LPF | Low-Pass Filter |
| NPV | Net Present Value (€) |
| NREL | National Renewable Energy Laboratory |
| OBU | Battery Use |
| PV | Solar photovoltaic |
| RES | Renewable Energy Sources |
| SAM | System Advisor Model |
| SCM | Self-Consumption Maximisation |
| SCR | Self-Consumption Ratio |
| SOC | State Of Charge (%) |
| SPB | Simple Payback (years) |
| SSR | Self-Sufficiency Ratio |
| TBU | To Battery Use |
| TGU | To Grid Use |
| TLCC | Total Life Cycle Cost (€) |
| TSO | Transmission System Operator |
| UÉvora | University of Évora |
| VAT | Value-Added Tax |
| VRE | Variable Renewable Energy |
| VRFB | Vanadium Redox Flow Battery |
| VRLA | Valve-regulated lead-acid |

## 1. Introduction

In 2020, the global additional installed power of battery energy storage systems (BESS) reached 5 GW, a 50% increase compared to the previous year [1]. Despite this significant installed power capacity increase, the role of BESS in the power grid remains vague: European Clean Energy Package (CEP), published by the European Commission in October 2019 [2] distinguishes storage from generation, transmission or load, to prevent double taxes when operating a battery (charge/ discharge), yet there is slow progress in the establishment of BESS development and use regulations.

When used at the energy system level, BESS can be conceived to provide different services, such as price arbitrage, capacity credit, ancillary resources (e.g. for voltage/ frequency regulation), integration of non-dispatchable renewable electricity production or smart charging for electric mobility [3]. At smaller scales, in residential, industry or services buildings, BESS can provide both maximised PV self-consumption and overall electricity reduction costs [4]. Through battery energy and power management, a smart control strategy can favour BESS performance in the execution of grid services such as peak shaving, curtailment avoidance or balancing of loads. Establishing an optimal energetic, technical and economical combination of performance goals is a complex task, highly dependent on manifold factors, such as, e.g. generation and consumption profiles, market regulation, applied electricity tariffs, grid connection parameters or system costs.

When considering more than one BESS unit, the optimal management strategy can be described as the delivery of multiple services or the simultaneous improvement of more than one objective. Although the cost-benefit assessment of such a system might be challenging given the lack of regulation [5], if several service revenue streams are "stacked", the investment in a BESS can be profitable [6].

Among the commercially available BESS technologies, the lithium-ion battery (LIB) continues to be the most widely used [1] in view of its cost decrease over the last few years. In addition to LIB, alternative chemistry batteries with potential cost decrease in the nearest future are the ones based on sodium and redox flow.

Given this context, this work investigates whether the hybridisation of LIB and VRFB technologies improves the competitiveness of an overall generation-storage-consumption system, addressing the following questions:
1. Does the LIB+VRFB HESS configuration improve the competitiveness of single VRFB or single LIB BESS configurations?
2. Which EMSs/ power allocation can be applied to HESS?
3. Is improving or maintaining the self-consumption rate possible, considering a seasonality factor in the battery state of charge operational range?

Aiming at raising awareness of the HESS management possibilities and associated technical aspects, this work is structured as follows: Section 2 delivers a literature review on HESS, Section 3 presents the overall methodology used in this work, Section 4 provides the main simulation results, and Section 5 addresses its assessment, with a final remark on future work (5.1). Finally, the fundamental conclusions of this work are detailed in Section 6.

## 2. Literature review of HESS

For the case of energy storage technologies, hybridisation is applied to any project that combines different energy storage technologies, generation, or load control technologies, co-located physically or virtually in a single network. Each BESS is combined to complement costs, performance, and environmental factors [7]. The HESS applied to microgrids with renewable energy sources (RES) is distinguished based on the application, capacity sizing, topology, configuration, energy management and control system [8].

Next, the HESS topologies, control techniques, and real-life demonstrators are summarised, and the HESS configuration studied in this work is justified.

### 2.1. Topologies

The integration of the HESS converter is selected considering the characteristics of cost, efficiency, controllability, complexity, and flexibility. Over literature, the HESS converters are categorised by the type of control, namely passive, semi-active or active [8].

Passive control is a simple and cost-effective solution where the different ESS units are physically connected, though they are not controllable. In the semi-active control, the inverter is connected to one ESS, while the other ESS is connected to the DC bus. This type of control is limited; however, the cost is lower than the active control. In the active control, each ESS is connected to its respective power converter. It is the most efficient and reliable converter even though the cost and complexity are higher when compared to previously mentioned topologies [8] [9].

## 2.2. Control, management, and power allocation strategies

HESS management and control are generally distinguished from classical and intelligent-based methods [8]–[10]. Classical methods are based on the ESS mathematical modelling and are suitable for real-time application, given the less computational effort; intelligent-based methods aim to maximise goals through optimisation functions, though with higher computational effort. The details of existing control methods are enunciated in Figure 1.

| HESS CONTROL | | | | | |
|---|---|---|---|---|---|
| **Classical control** | | | | | |
| | Filtration-based | Rule-based | Dead-beat | Droop-based control | Sliding mode |
| | • High filter and low filtering methods.<br>• Efficient for the control of charge/discharge conditions.<br>• Less complex and well suited for real-time control. | • Decision making process pertaining to the control objective.<br>• Less computational effort, easy and simple to implement. | • Model based.<br>• Generates the ratio of duty cycle to minimize error regulation in one control cycle.<br>• Simple implementation and easier process involvement. | • Highly reliable, decentralized, and easy implementation. | • Non-linear control which toggles between the control laws based on the state vector.<br>• Robust. |
| **Intelligent-based control** | | | | | |
| | Model predictive controller | Neural network (ANN) and fuzzy logic (FL) | Optimization based | Unified controller | |
| | • Prediction of the future behaviour of the system.<br>• It provides a uniform approach to its design, with easy incorporation of constraints.<br>• Possibility to control high number of controls variables. | • FL controllers are easier to implement and less sensitive to parameters' change.<br>• Modelling isn't required.<br>• ANN is a mathematical model which was developed to recognize and process parallel data, consisting of several machine neuron layers. | • Linear programming (LP), Dynamic programming (DP), genetic algorithm (GA);<br>• Multi objective and Evolutionary - On one hand there are the evolutionary methods, that can handle with multiple objective functions with better response than conventional; on the other there are the multi-objective which can be controlled at a time. | • Faster dynamic voltage regulation, effective power sharing under any sort of disturbances, reducing fluctuations in rate of charge/discharge of battery, and power quality enhancement.<br>• Feed forward to manage the power flow between batteries; produces switching times and sequences of the state vector.<br>• Higher performance than conventional controllers and less sensitive to parameters change. | |

Figure 1. The HESS control methods summary. Based on [8]–[10].

Relevant works on the field of energy management of energy storage and energy sources hybridisation are found in the literature but mostly for general hybrid energy systems, without a clear focus on the hybrid energy storage system, and in this case, do not approach the different ESS technologies interaction as the present work does. In [11] the authors optimise the integration of different batteries and different generation sources (PV and wind) through the use of valuable parameters such as the LCE (life cycle emission) and COE (cost of energy), although integrating single ESSs configurations. Also, the authors of [12] present a study on the predictive energy management strategy based on machine learning applied to buildings, and the authors of [13], where a novel two-step approach to deal with hybrid energy sources to optimise the techno-economic aspects, sizing the system and minimising the LCOE, considering operation and management. The authors of [14] study the optimal operation management of a microgrid with two distinct objectives (operation cost and emission propagation), considering an optimal solution based on fuzzy logic approach.

With regards to the hybridisation of batteries (in this work approached as electrochemical and electrical devices), approached in this work as HESS, the following state of the art is described. The HESS potential has been identified for several energy-consuming sectors. For transports, the combination of batteries and supercapacitors is the most studied configuration, as referred to in work developed in [17], which considers a power allocation strategy based on a low-pass filter (LPF) and fuzzy logic (FL) control technique. Within the transport sector, lithium-ion batteries with different cathode chemistries, LFP and LTO, are addressed in [18] and further validated, demonstrating a more significant lifetime of hybridised LFP technology rather than single LFP technology. Fuel cells and lithium-ion batteries are investigated in [19], considering online and offline EMS methods to improve fuel consumption and source lifetime. The authors of [12] investigate the interaction of fuel cells, lithium-ion batteries and ultracapacitors for HEV application, which are sized as an optimisation problem. The authors of [15] address fast and short-term fluctuations of PV systems at the residential scale, including impact analysis on the battery-supercapacitor HESS group. The work focused on presenting the influence of dynamic battery operation influence in the studied KPIs, considering the optimised HESS sizing. In [16], the authors developed a techno-economic optimisation for the size and power management of a residential use-case (PV, battery, EV charger, load consumption) to optimise PV self-consumption, day-ahead and frequency containment reserve (FCR), also account for dynamic battery degradation.

A fuel cell and a nickel sodium chloride battery are studied in [20] using electrical battery models, and real-time implementation is achieved. Economic approaches are also addressed, such as in [21], where six configurations of batteries and flywheels NPV and LCOE indicators were calculated for a Greek island application and compared with single ESS scenarios. The configuration of VRFB and LIB is investigated in [22] to improve the BESS lifetime. Four scenarios are studied, with the calculation of a few energy indicators (to and from the grid). With this configuration, authors in [23] studied the HESS integration with RES, while an EMS uses a predetermined forecasted and scheduled power curve to evaluate power mismatch. Though the hybridisation of batteries is being addressed in the literature, the topic is still recent, and studies reveal a lack of complete assessment which could fairly address real-time EMSs, considering the interaction of batteries.

*2.2.1. Demonstrators*

Project commissioning validates the HESS power control allocation challenges in real-time suitability, performance, and KPIs analysis. Several HESS worldwide demonstrators can be found:

- HESS RES integration and isolated application; validation of configurations and models; technical optimisation (e.g., battery degradation, thermal stress, performance):
    - 2021, HYBRIS [24] with HESS (1 MW /0.47 MWh) demonstrators in Italy, Belgium, and The Netherlands.
    - 2020 [25] investigates high-output lithium-ion batteries with high-capacity lead-acid storage batteries to facilitate wind generation systems applied in Poland.
    - 2016, The Duke Energy Rankin Project [26] combined a battery and supercapacitor, 100 kW/ 300 kWh, installed in Gaston, North Caroline, USA.

- HESS simulation and management for control and stability purposes:
    - 2020, HyFlow [27] with a high-power vanadium redox flow battery (RFB) and a supercapacitor, advanced converter topologies and a flexible control system.
    - 2019, the SHAD project [28] developed AC and DC solutions to control different battery technologies: supercapacitors, flywheels, or fuel cells, for multiple applications.

- HESS grid services and integration – Algorithms, performance, suitability:
    - 2019 [29] studies a 2 MW/ 5 MWh RFB and LIB with 48MW/ 50MWh for grid connection in the UK.
    - 2018 [30], with a 4 MW/ 20 MWh liquid sodium-sulphur battery (NaS) and 7.5 MW/ 2.5 MWh of lithium-ion. Applied in 17 districts and four cities in Niedersachsen, Germany.
    - 2017 [31], with 300kW/ 1MWh flow batteries to hybridise with a LIB of 120 kW, Melbourne, Australia.
    - 2016 M5BAT project [32]. Study of lead batteries: one flooded and one VRLA, 1.3 MWh OCSM battery system, two strings of 1 MWh VRLA gel technology and three lithium-ion technologies.

- HESS for EV application – Fast charging and models development:
    - 2020, Hydealist [33] combines supercapacitors and conventional batteries.

- 2019, Hybrid Battery Optimisation [34] combines lithium-ion batteries and supercapacitor to optimise energy and power.

**2.3. Review of the hybridised batteries technologies**

A BESS is defined through different performance specifications: efficiency, response time, power and energy density and capacity, ageing, and degradation effects. The interconnection of different BESS can result in a solution with complementary characteristics, resulting in an overall increase in the performance of the system [22]. Selecting a specific type of storage for an application arises from technical and economic optimisation.

This work addresses different control combinations of commercially available LIB and VRFB technologies aiming to design a HESS due to its distinct energy and power characteristics. The technologies' specifications were previously documented [35] [36] [37] and are summarised in Table 1.

Table 1. Typical characteristics of VRFB and LIB [35] [36] [37].

| ESS Technology | Vanadium Redox Flow Battery (VRFB) | Lithium-Ion Battery (LIB) |
| --- | --- | --- |
| Power range (MW) | 0.03-3 | Up to 0.01 |
| Typical discharge time | sec-10h | min-hour |
| Overall power efficiency | 0.65-0.85 | 0.85-0.95 |
| Power density (W/L) | ~<2 | 1500-10000 |
| Energy density (Wh/kg) | 10-30 | 75-400 |
| Storage durability | hour-months | min-days |
| Self-discharge (per day) | Small | 0.1-0.3 % |
| Lifetime (year) | 5-10 | 5-15 |
| Life cycles (cycles) | 10000-13000 | 1500–4500 |

VRFBs are suited for applications that require several hours of storage, such as peak shaving, spinning reserve, stabilisation, and dispatch, mostly for stationary or mini/microgrid applications. Its main components are pumps, storage tanks, a stack composed of cells, piping, a control unit and power conversion equipment [38]. RFB technology is tolerant to over-discharge and overcharge avoidance is achieved through the control of the reference open-circuit voltage outside the stack. Vanadium ions are present in both electrolyte tanks, and crossing through the stack membrane does not lead to contamination of the electrolyte (crossover effect), which occurs in other RFB technologies, e.g., zinc-bromine.

The life expectancy of the VRFB is 10-15 years, corresponding to about 1000 annual cycles, though it can last more than 20 years [38]. VRFB have virtually no degradation (0-100 % of usable SOC) [39]. Some batteries have an inert gas on top of the electrolyte to avoid possible side reactions with the oxygen of the air, and in the case of the University of Évora VRFB, argon gas is used. VRFB cycle efficiency is in the range of 75-85 %. Its scale-up can be achieved at any power/energy rating (independently), with capacity being customisable with the sizing of electrolyte tanks and power sizing through the sizing of the stack. Vanadium's environmental restrictions are less stringent than those for lead and cadmium. Powered vanadium metal is combustible, while most of the vanadium compounds are not, and, in general, do not constitute a fire or explosion hazard [38], which stands out as an advantage of RFBs in contrast with other battery chemistries such as LIB.

LIB has a high energy density [35], making them suited for short-term and medium-term applications, such as frequency regulation, voltage support, or peak shaving [40]. Generally, it has two electrodes and an organic, non-aqueous electrolyte containing dissolved lithium salts with efficiencies between 80-90 %, with case studies of up to 97% [35] [41]. Typically presents an average life cycle of 5-15 years [41]. Its power and energy are easily scalable, and its production costs are expected to keep decreasing in the near future due to large-scale manufacturing [40].

Typically, an active cooling system is integrated to avoid extreme temperature ranges, to avoid enhanced energy storage capacity degradation. Electronic voltage control limits are crucial to prevent cell damage, SOC determination and cell operation on-stop [41].

When compared to other BESS, LIB components are less toxic, however, they are still an environmental hazard when not disposed of appropriately, which is the case in some countries where they are disposed of in landfills [42]. Given its good state after high-intensity applications, LIBs could be re-used in lower-intensity applications as grid storage or serve off-grid applications, which adds a stage to the LIB life cycle before the end of life (EOL). Recycling is crucial for sustainable technology evolution. However, currently, its recycling rate lies below 3% worldwide.

## 3. Methodology

A complete technical-economic model is developed using MATLAB software to discuss the research questions referred to in Section 1. The model reproduces the operation behaviour of BESSs and other microgrid equipment, computes the EMS/power allocation algorithms and delivers the evaluation of the overall system based on the calculation of key performance indicators (KPIs) from an energetic and economic perspective. Over 15 years of operation are analysed, with the algorithms running on a 1-minute basis. The developed approach allows the BESS to be operated either in a single or hybrid scenario and the input data to be easily inserted. Moreover, the model addresses the combined advantages and drawbacks of the LIB and VRFB joint control and the possibility of delivering multiple tasks and still being competitive.

The following subsections describe key aspects of the model inputs and algorithm-based computation of the developed LIB and VRFB scenarios.

### 3.1. Microgrid description

The microgrid of the Renewable Energies Chair of the University of Évora allows the operation monitoring of ESSs and PV installations, controls deployment, testing, and model development. The microgrid integrates two ESSs, a 5 kW/60 kWh VRFB from redT [43] and a 5.0 kW/9.8 kWh LIB RESU10 battery pack from LG [44]. It also accounts for two PV installations, one with 6.7 kWp crystalline PV installation and the other with a 3.2 kWp amorphous BAPV system. In this work, the HESS configuration achieves an 8-kW nominal power capacity (LIB inverter limits output power) and a 70-kWh energy capacity as a result of combining the VRFB and LIB technologies referred in Section 2.3. The microgrid layout is shown in Figure 2.

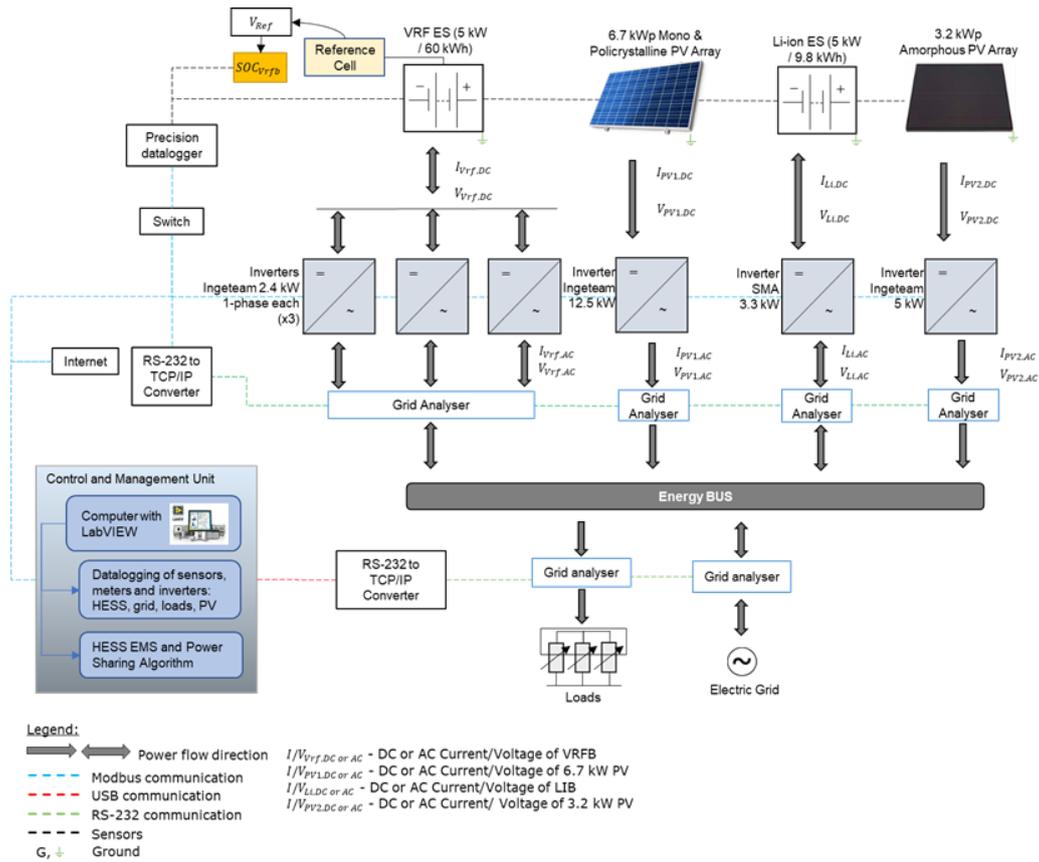

Figure 2. HESS microgrid architecture.

The PV generation power capacity is close to 10 kW, with 6.7 kWp of monocrystalline and polycrystalline technology and 3.2 kWp of amorphous technologies; both installations are connected to individual inverters from Ingeteam. The ESSs are composed by a 5kW/ 9.8 kWh LIB connected to a 3.3 kW SMA inverter and a 5kW/ 60 kWh VRFB connected to a total of 7.2 kW monophasic inverters from Ingeteam. Each technology is connected to AC energy analysers, temperature sensors and DC measurement devices. The microgrid control is achieved by an in-house developed LabVIEW [45] program, based on Modbus TCP/IP protocol. It includes real-time data logging of energy analysers and sensors, the EMSs and power-sharing algorithms.

### 3.2. PV installations and consumption inputs

The PV generation profile results from 2019 data treatment of the two existing PV systems. The data acquisition of both PV systems is made at rates of 1s and 2s. The data sets were treated separately, averaged to a 1-min period and later used as input for the modelling. 1-minute data averages were chosen to shorten the simulation running time whilst still taking into account PV generation fluctuations.

The consumption profile used is a 15-minute average of the data made available by the Portuguese DSO company E-REDES, for 2019, for each Portuguese energy consumption sector for the normal low voltage (NLV) buildings sector [46]. The consumption load profile data was processed to correspond to the PV generation 1-minute time resolution. Figure 3 shows one-week examples of PV generation and consumption profiles for (a) winter and (b) summer.

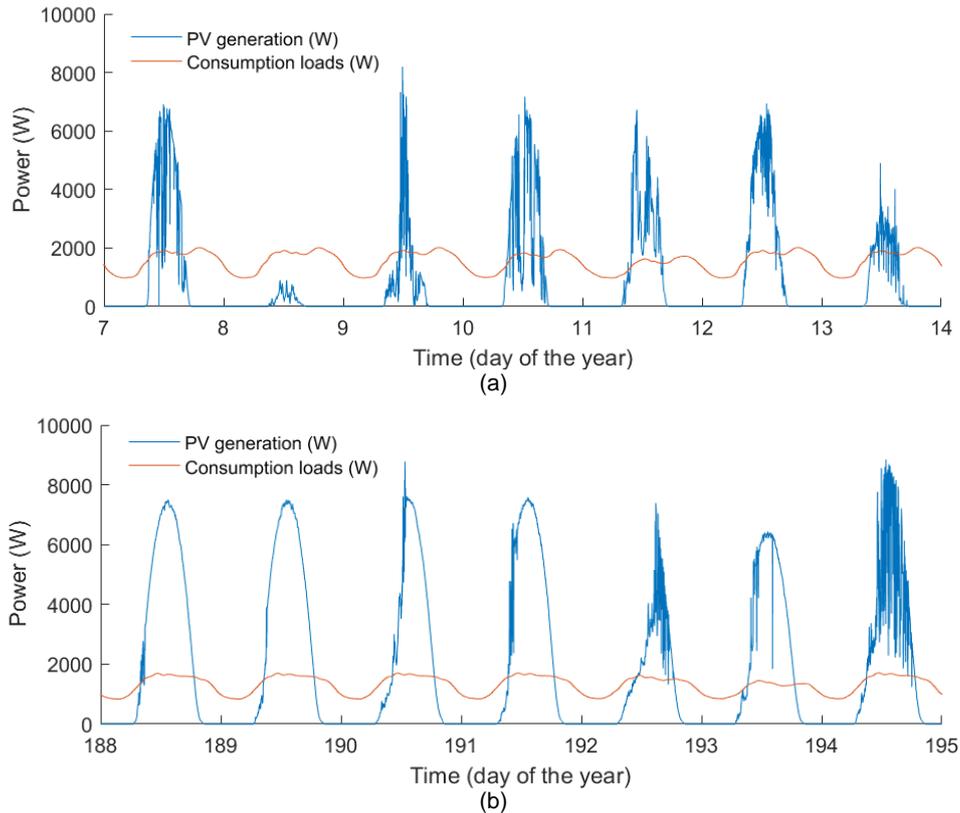

Figure 3. PV generation and consumption data used as input, with one-week examples for (a) of winter and (b) summer. The PV generation profile is a result of the 1-minute data treatment of the two PV systems (3.2 and 6.7 kWp) and the consumption profile obtained.

Considering the 1-minute timeframe, the PV generation and consumption profiles were used for a 15-year timeframe, considering average yearly PV degradation and considering a predicted increase rate of electric energy consumption in the buildings sector, as detailed in the following sections.

### 3.2.1. PV degradation

The degradation performance of PV installations is a worldwide research topic to assess lifetime of PV systems better and provide information to the manufacturers to work on improving its modules lifetime. Generally, the modules with no relevant defects over their lifetime are in agreeance with the manufacturer's warranty not degrading more than 20% of their initial power in the first 25 years of operation (-0.8%/year). In ordinary situations, the degradation value is even lower. A study of 110MW of PV installations, in 25 installations over Europe with installations operation data records up to 30 years, showed PV modules degradation ranging from -0.1%/year and -0.75%/year, respectively, from different manufacturers [47].

In the context of this work, a PV module annual efficiency degradation of -0.45%/year was considered as reasonable and was applied to the input PV data along the simulation timeframe.

### 3.2.2. Consumption profile increase

Future energy consumption projections for 2030 and 2040 were presented in a study developed by the International Energy Agency (IEA) [48], where three main scenarios were shown, regarding the "stated policies", the "sustainable development", and the "current policies" for the years 2030 and 2040, to be compared with actual data of the 2000 and 2018 years. The global electricity demand by scenario for the building sector is shown in Figure 4 [48].

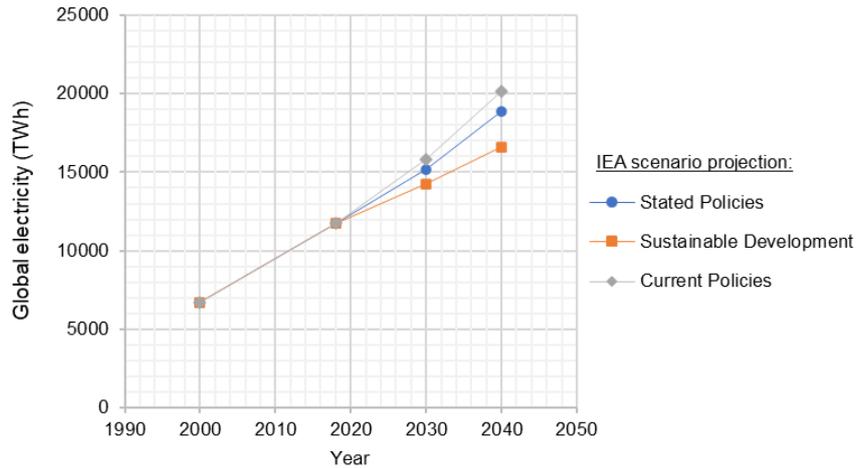

Figure 4. Global electricity demand for the building sector, in TWh [48].

Considering the proposed IEA future energy demand scenarios for 2030 and 2040 observed in Figure 5, and the lack of detail on how these projections could impact each building profile (in an undifferentiated way, e.g., services, residential), a 2 %/year increase was considered in the consumption profile mentioned in Section 3.2, to scale-up future electric consumption and complement the analysis of this work. Besides the associated uncertainty in IEA projections, the inclusion of this energy demand projected increase rate can lead to a better approximation of future electricity energy consumption in buildings.

### 3.3. Battery models

The description of each BESS operation is obtained through its simulation model. The batteries' physical constraints must be included in the model as boundaries, such as the maximum allowable power and capacity, SOC and depth of discharge (DOD) limits, efficiencies and lifetime. Other constraints are related to the model's initial definitions and are battery-related as well (e.g., initial SOC).

The replacement of the battery at a specified capacity should be defined at the beginning of the project. In this work, the authors considered the battery replacement in the 15$^{th}$ year, the final year of the analysis period.

#### 3.3.1. VRFB modelling

One of the modelling approaches used to describe the multiphysics of the VRFB is the nonlinear Nernst-Plank equation which is included in an equivalent circuit that describes the significant reaction occurring in the real-time operation of VRFB with reduced complexity and satisfying results. The VRFB was firstly fully experimentally characterised, and the model was validated with an energy management strategy developed in [49]. The model includes thermal effects, transients, dynamic SOC and auxiliary power consumption.

In this work, the available defined SOC ranges from 5 to 95 %, with the EMS initiation at 50% SOC. The available charge and discharge power range is from -5000 to 5000 W and considers a power-SOC relation. The battery is operated at 20-35°C, and the energy capacity degradation is 0 %/year [39] [50].

Regarding the previously calculated efficiency of the VRFB-connected power electronics, its inclusion is made, and the inverters' standby mode is considered to be 30 W.

*3.3.2. LIB modelling and lifetime*

The authors used real-time battery operation and data acquisition to develop a LIB model for the referred battery based on the equivalent electrical model and the modified Shepperd model [51]. This model did not consider battery ageing since it was based on data obtained at certain current (power) levels.

LIB ageing strongly depends on operation conditions: temperature, SOC, total energy throughput (electrochemical operating windows), and charge and discharge operating rates. Consensus exists on the lack of a universal lithium-ion lifetime model, although calendar life and the number of cycles are recognised as fade mechanisms [52]. The charge and discharge cycles' patterns and the operating conditions can further decrease lifetime. The battery energy capacity reduction due to calendar and charge-discharge cycles and lifetime fade determines the replacement of battery systems, impacting the overall investment plan. The voltage curve is dependent on the battery SOC, the operating current, resistance, and energy capacity. The internal resistance increases with the calendar time, causing a decrease in the usable voltage range. The operation of the battery at higher currents implies a higher speed voltage decrease, reducing the use of the battery bulk capacity. The thermal behaviour of the model is generally obtained using a heat transfer with the environment to represent the instantaneous thermal effects (energy losses), affecting the battery energy capacity and the internal resistance.

In this work, the LIB model is based on the combination of the model developed in [51] and the lifetime prediction model developed by the National Renewable Energy Laboratory (NREL), for the NMC LIB technology, included in their System Advisor Model (SAM) software. SAM is dedicated to commercial battery systems suited for real-time battery control algorithms [52]. The energy capacity loss is described in calendar time or in cycle number, and in this work, the authors follow the calendar time approach. To describe the voltage reduction with calendar time, a constant internal resistance is attributed to represent the impact on voltage.

The NREL lithium-ion ageing model uses Eq. (1) and (2) to determine the adequate battery energy capacity (%) [52],

$$q = q0 - kcal \times \sqrt{t} \qquad (1)$$

Where $q0$ is given as 1.02 fraction, $t$ is the day, and the stress factor, $kcal$, is given in the following Eq. (2),

$$kcal = a \times e^{b\left(\frac{1}{T} - \frac{1}{296}\right)} \times e^{c\left(\frac{SOC}{T} - \frac{1}{296}\right)} \qquad (2)$$

Where the coefficient $a$ is 0.00266 in unit of $1/\sqrt{day}$, $b$ is -7280 K, $c$ is 930 K, and T is the temperature in K.

Regarding battery voltage, the internal cell resistance considered is 0.001155 Ω (extracted value from SAM). Concerning the battery's thermal behaviour, the effective capacity varies with the ambient temperature, being 100 % at 23 °C, which is considered to be a reasonable value given the controlled environment provided by the air conditioning unit.

The LIB operating SOC range is 10-90%, and the SOC of 50% was chosen for its initial state. The inverter connected to the LIB has a nominal power capacity of 3.3 kW, which is assumed as the maximum power capacity retrieved from the battery. In that sense, the inverter limits for charge and discharge power range are from -3300 to 3300 W. For the standby mode of the LIB inverter, a value of 5 W is used.

**3.4. Energy Management Strategies: Different HESS Power Allocation Scenarios**

The energy management strategy baseline is the self-consumption maximisation strategy, considering the following premise: the energy fluxes priorities are the same for the different scenarios. Here, the PV generation is primarily self-consumed. The BESS only charge with the surplus PV-generated energy and discharges to supply consumption loads, while the grid is seen as the last resource to exchange energy with the building grid. The subjacent EMS prioritises the PV generated energy for direct self-consumption and, if not possible, the exceeding generated energy is sent to charge the batteries; otherwise, it is sent to the grid. From the prosumer perspective, the priority levels are the consumption of PV generated energy and then from the battery. If neither of those possibilities is available, the energy is consumed from the grid. A simple description of the algorithm architecture is presented in Figure 5.

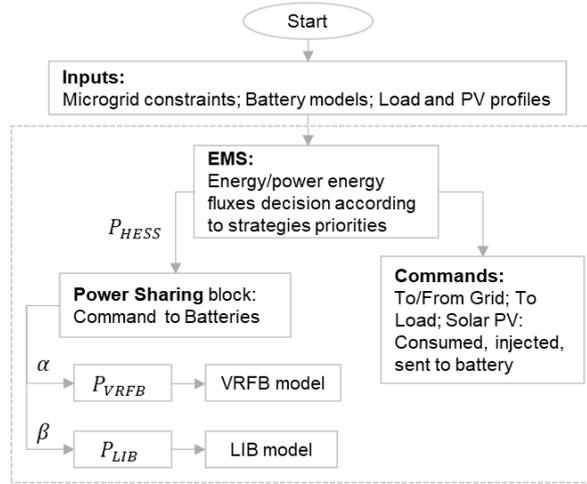

Figure 5. EMS and power allocation strategy for the HESS case study.

To properly analyse the EMS and power allocation control, which defines the operation algorithms of the HESS, five scenarios have been built, including the previously referred inputs. Scenarios 1, 2 and 3 obey a HESS power allocation method, consisting in a power filter applied to the overall HESS power command, expressed in equation Eq. (3):

$$P_{HESS} = \alpha P_{VRFB} + \beta P_{LIB} \tag{3}$$

Where $\alpha$ and $\beta$ are constants that establish a percentage of the total battery power command, $P_{HESS}$. The $P_{VRFB}$ and $P_{LIB}$ are the VRFB and the LIB power commands, respectively. In the following, the five scenarios are described.

- *Scenario 1:* Consists in the roughest HESS power command allocation. The power allocation among the HESS is defined as constant percentage values of the total power command. The battery with the greater energy capacity will be attributed with the highest percentage of the overall battery command, and in this case, the vanadium redox flow battery. In that sense, and considering Eq. (3), it includes a constant value of $\alpha = 0.75$ and of $\beta = 0.25$ for LIB in their defined SOC range.

- *Scenario 2:* Aims at using the energy capacity of the LIB in a controlled way to increase its lifetime (relying on the LIB P(SOC) relationship to avoid extreme levels of power and SOC). The LIB P(SOC) relation is detailed in Figure 6 (a) and (b). In this case, the VRFB sustains the baseload of the consumption profile.
  The variable $\beta$ is dynamic, and as such, at each step is updated, as well as the LIB power command. The constant $\alpha$ is obtained by $\beta - 1$ through the subtraction of one unit. Scenario 2 is based on the P(SOC) relation defined for the LIB, based on the obtained experimental data. This definition is explained through Figure 6.

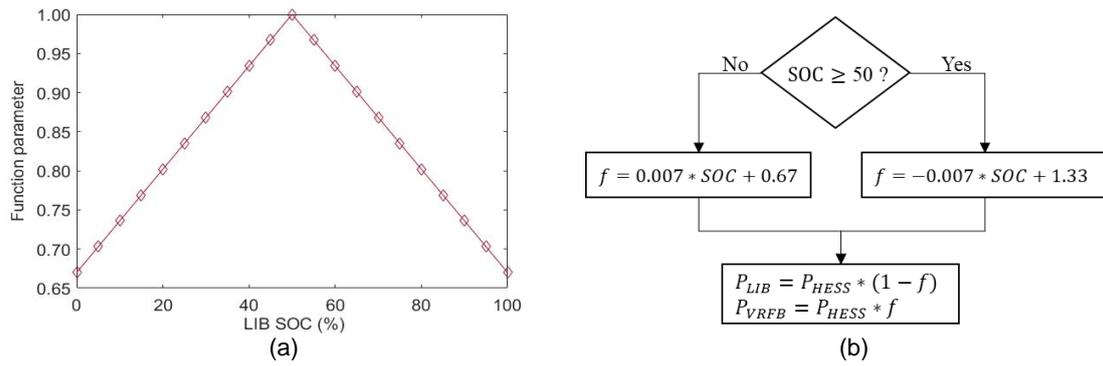

Figure 6. Details of Scenario 2 of LIB operation based on (a) the function with the (b) P(SOC) equations flux gram detailed, where $f$ is the intermediate calculus power command.

- *Scenario 3:* This scenario aims to be an algorithm capable of operating both batteries, including the execution of distinct EMS independently. LIB executes a peak-shaving approach, with its operation among a power range above the -1000 W and below the 1000 W, while VRFB power command fulfils, within its possibilities, the rest of the needed energy/power.

- *Scenario 4:* Given the seasonal PV variation and consumption profiles, this scenario intends to analyse the impact of not choosing the maximum possible SOC operation range of the batteries. The impact is evaluated by observing three indicators: SCR, LCOE and OBU (defined further ahead).
  Different battery SOCs ranges are investigated, considering two cases: summer and winter. In the first, the SOC range is made variable in summer and constant in winter; in the second, the opposite is studied. Additionally, a third case presents the optimum configuration for each indicator.
  Scenario 4 arose in the context of the noticeable effects in the results of Scenarios 1-3, given that the batteries SOC lay in extreme limits during winter and summer. In the winter, it lays at lower SOC values ; therefore, less available discharge capacity exists during this period. In the summer, it lays at higher values and therefore, less available charge energy capacity exists. Scenario 4 aims to reduce those effects, and it analyses the impact of not choosing the maximum possible SOC operation range of the BESS. For that, the SOC ranges for both battery technologies were combined and simulated, as expressed in **Table 2**.

**Table 2.** LIB and VRFB SOC range definition studied on the base of Scenario 4.

| LIB Minimum SOC value | 10, 20, 30 or 40 |
| --- | --- |
| LIB Maximum SOC value | 50, 60, 70, 80 or 90 |
| VRFB Minimum SOC value | 5, 15, 25, 35 or 45 |
| VRFB Maximum SOC value | 55, 65, 75, 85 or 95 |

To avoid deceptive results, a lower DOD limit is needed for the values referred in Table 2. DODs lower than 40 % were not considered output results since an investment in those would not be justified (for example, 40-50% SOC range).

- *Scenario 5:* This scenario provides baseline scenarios, with a single battery case for VRFB and a single battery case for LIB, i.e., the use of one battery to execute the EMS of self-consumption maximisation.

The batteries SOC range are defined in Section 3.3, except for Scenario 4, due to the seasonal SOC range variation.

### 3.5. Evaluation indicators

An appropriate evaluation can be obtained through the determination of performance indicators related to investment and energetic perspectives.

As the most relevant economic indicators, the analysis relies on the determination of the following indicators: the Net Present Value (NPV), expressed in €; the Levelized Cost of Energy (LCOE), expressed in €/kWh and is based on the calculation of the Total Life-Cycle Cost (TLCC) and in the Present Value of the Operation and Maintenance Cost (PVOM); the Internal Rate of Return (IRR) expressed in %; and on the Simple Payback (SPB), expressed in years. The indicators are calculated based on the equations expressed in [53], explained in detail in previous works of the authors, as in [54].

The chosen energy indicators are determined for one full year, and the presented result is the average of the obtained annual values. The analysis comprises an energetic evaluation with the following indicators: self-consumption ratio (SCR), self-sufficiency ratio (SSR), grid-relief factor (GRF), battery use (OBU), energy from the grid (EG), from grid use (FGU), to grid use (TGU), from battery use (FBU) to battery use (TBU). These energy performance indicators are already detailed in previous authors' works, as in [55] or some in [56].

### 3.6. Economic analysis and tariffs details, and related inputs

The economic analysis considers the effectively consumed solar PV energy, the discharged energy from the battery or batteries (depending on the scenario), and the energy sent to the grid. The expenses considered are concerning to the energy that is received from the grid.

The economic inputs which compose the different scenarios are detailed in this section. Table 3 presents the overall economic inputs and tariff details used.

Table 3. Economic input expenses for modelling tool

| Equipment/component | Value (with VAT) |
|---|---|
| Module price (€/Wp) | 0.35 * |
| Total PV installed power modules (Wp) | 9750 (30 modules of 325 W each) |
| Discount rate** (%) | 10 [11] |
| Inflation rate (%) | 7.2 [57] [11] (average) |
| Annual Energy increase rate (%) | 5.6 [57] (average) |
| 6.7 kWp BAPV inverter (€) | 2432 [58] |
| 3.2 kWp fixed-mounted PV inverter (€) | 1483 [59] |
| Lithium-ion battery (€) | 4527 [60] |
| Lithium-ion battery inverter (€) | 2125 [61] with VAT and delivery to Europe |
| VRFB (€) | 48080 (2404 €/kW + 601 €/kW; € ≅ USD) [62] |
| VRFB inverters (€) | 3×1159 [63] |
| Cabling, installation, PV structure and related (€) | 2850 * |
| Overall system operation & maintenance (OPEX) (€/year) | 500 *** |
| Daily contracted power tariff (€/day) | 0.2796 [64] |
| High/ low tariff (€/kWh) (bi-hourly) | 0.2116/ 0.1145 [64] |

* Company quotes from 2023; ** Depends on expectation; *** VRFB tank inert gas (highly pure argon).

Some expenses presented in Table 3 are eliminated for Scenario 5, given the single ESS instead of the HESS. In the case of Scenario 5-LIB, the following costs are considered null: VRFB, VRFB inverters and OPEX related to VRFB. Regarding Scenario 5-VRFB, the LIB and its inverter costs are considered null. The remaining costs are not altered, except for the cost of cabling, installation, PV structure and related, which are reduced by one-third of the presented value in Table 3 to better to fit the scope of the single ESS analysis.

Inflation and energy rates were extracted from PORDATA website, which presents the updated and yearly values of Portuguese statistics. The used values correspond to the 2022 year. The discount rate consulted to be in the range of the published works [11] [65].

Benefiting from the bi-hourly tariff, the considered energy value cost (€/kWh) depends on the consumption period. A contracted power of up to 6.9kVA and 100 kWh/month of energy consumption was considered [64].

## 4. Results

This section addresses the main outputs of the study, considering the developed simulation tool used to evaluate the LIB+VRFB HESS configurations and the single ESS scenarios detailed in Section 3.4. Following the proposed methodology, the HESS and single ESS scenarios 1-5 were built to answer the research questions enunciated in Section 1.

Scenario 1 is the first approach to HESS operation and is considered a direct method that makes a soft approach to power distribution, considering only the energy capacity of the batteries. Scenario 2 presents an approach to better control the LIB operating conditions and prevent operation in limit boundaries through obeying the P(SOC) real-time relation (SOC measurement dependent). Scenario 3 considers the use of LIB to fulfil the values within the range of -1000W to 1000W, identified as a similar approach to peak-shaving using an ESS. Scenario 4 evaluates the seasonal variation of the SOC range of the batteries. With scenario 4, it is possible to vary the batteries' SOC range to favour the result based on the seasonality of PV generation and consumption profiles. Scenario 5 is used as the baseline, with configurations for the single BESS.

The five scenarios will be evaluated through KPIs in the techno-economic assessment.

### 4.1. Scenarios 1, 2, 3 and 5

The results of the different Scenarios defined in Section 3.4., excluding Scenario 4, are presented through the KPIs obtained in Figure 7, below. The ideal best value of each energy indicator is represented with an orange line to improve the readability of Figure 7. The best value is attributed from the point of view of the prosumer, previously detailed in Section 3.4. The prosumer prioritises the PV self-consumption and aims to reduce the use of the batteries and the grid, simultaneously reducing the use of the batteries and the grid.

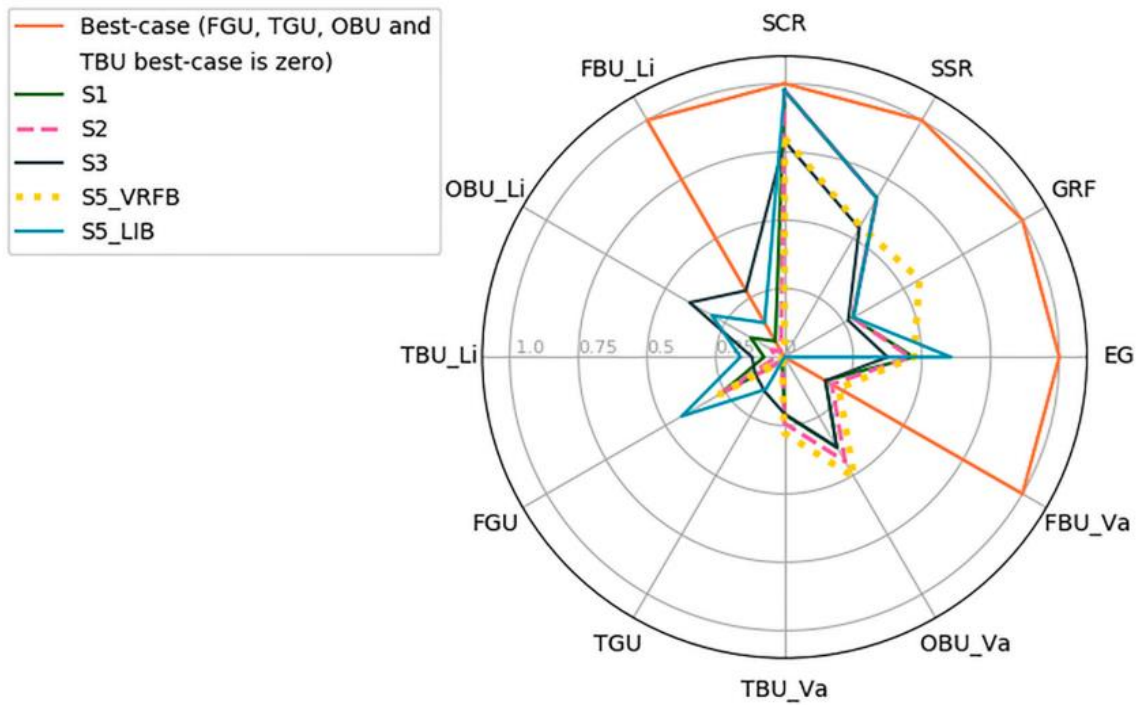

Figure 7. Battery-related energy KPIs for Scenario 1, 2, 3 and 5. EMS-related energy KPIs for Scenario 1, 2, 3 and 5. The "_Va" corresponds to the VRFB and "_Li" to the LIB output values. Each scenario is represented as "S".

The economic indicators were calculated for Scenarios 1, 2, 3 and 5. The economic KPIs NPV, LCOE, IRR and SPB are presented in Table 4 for each case.

Table 4. Economic KPIs of Scenarios 1, 2, 3 and 5.

| Configuration | HESS | | | VRFB single | LIB single |
|---|---|---|---|---|---|
| Scenario/Economical parameter | Scenario 1 | Scenario 2 | Scenario 3 | Scenario 5 | |
| Total investment (€) | 68403 | 68403 | 68403 | 59851 | 14946 |
| NPV (€) | -4140 | -3821 | 1251 | 4768 | 24764 |
| LCOE (€/kWh) | 0.46 | 0.46 | 0.35 | 0.42 | 0.28 |
| IRR (%) | -3.01 | -3.00 | 3.00 | 2.00 | 31.0 |
| SPB (years) | 9.69 | 9.62 | 8.21 | 8.69 | 4.10 |

### 4.2. Scenario 4 results

The SOC ranges defined in **Table 2** are tested, for the EMS and power allocation of Scenario 1, given the simplicity of the approach and overall worst performance. To evaluate this scenario, three main KPIs are observed: the priority of the EMS is to maximise self-consumption ratio, and, in this case, SCR proves its relevance; the second priority is to minimise the battery utilisation, which can be validated with the OBU indicator; and, finally, the third priority is to observe the LCOE indicator, to address the financial return better. The winter and summer cases are compared with the baseline of all year variable SOC range (corresponding to Scenario 1 simulated by SOC combinations in Table 3).

As previously outlined, Scenario 4 results are based on data profiles and conditions defined along with the methodology and, for this reason, are valid only within the defined conditions. Table 5 presents the three best cases results of this optimisation, considering the output of the previously mentioned evaluating indicators.

Table 5. Scenario 4 output results. SCR obtained value is the priority of the overall Scenario, LCOE and OBU are the second priorities. Uc means "use-case".

| KPI | Uc1 - Variable throughout all year | | Uc2 - Winter variable, summer fixed | | Uc3 - Winter fixed, summer variable | |
|---|---|---|---|---|---|---|
| | Best value | SOC range | Best value | SOC range | Best value | SOC range |
| SCR | 0.848 | VRFB decisive: [5, 95] | 0.848 | VRFB decisive: [5, 95]; [15, 95]; [25, 95]; [35, 95] | 0.848 | VRFB decisive: [5, 65]; [5, 75]; [5, 85]; [5, 95] |
| LCOE (€/kWh) | 0.797 | LIB determinant: [10, 90] | 0.797 | VRFB: [5, 95]. LIB: [10, 60]; [10, 70]; [10, 80]; [10, 90] | 0.797 | LIB decisive: [10, 90] |
| OBU | LIB decisive: 0.110 | VRFB: [5, 95]; LIB: [40, 80] | LIB decisive: 0.143 | LIB decisive: [40, 80] | LIB decisive: 0.135 | LIB decisive: [40, 80] |

### 4.3. Discussion

Figure 7 shows that neither of the proposed scenarios corresponds to an optimum for all the calculated energy KPIs compared with the defined ideal best values. Scenarios 1, 2 and 5-VRFB present similar SCR, SSR, GRF, EG and FGU indicators. The highest VRFB energy exchange is obtained in Scenario 5 (OBU_Va), and for LIB in Scenario 3 (OBU_Li).

In Scenario 1 the VRFB is the primary energy exchanger (OBU, FBU and TBU). Table 4 shows that this configuration is the least attractive approach of all scenarios, considering the obtained economic KPIs of NPV, IRR and SPB.

In Scenario 2, the VRFB is a promising candidate to operate with LIB, given the different energy capacities and theoretical 100% DOD. From the energetic KPIs in Figure 7, the LIB use is highlighted as being the lowest compared to other scenarios. The economic KPIs results present a slightly better output, although close to scenario 1.

In Scenario 3, the LIB and VRFB batteries' energetic use are closer. Figure 7 shows the generally lower SCR, SSR, FGU, GRF than Scenarios 1, 2 and 5-VRFB, but higher in the ones associated with the LIB use (OBU_Li, FBU_Li, TBU_Li). Economically, Scenario 3 presents better financial results than the studied HESS approaches for all the KPIs studied. It also presents the second-lowest LCOE and second-lowest SPB and the second-highest IRR. Scenario 3 indicates that the HESS LIB+VRFB configuration with the suited EMS and power allocation method can improve the competitiveness for the VRFB single case, and although the highest investment, the batteries are better energetically managed.

In Scenario 4, with the help of Table 5, the priority is achieved with the best SCR maintained as 0.848 for the three cases. The second priority can be the LCOE (lowest cost) or the OBU (lowest battery energy exchange). Regarding Uc1 (variable all year), the lowest LCOE of 0.797 €/kWh corresponds to a VRFB SOC range of 5-95 % and 10-90 % for LIB. The lowest use of the batteries presents 11 % for LIB, maintaining the SCR, as the VRFB with 5-95 % and LIB with 40-80 %. Regarding Uc2 (winter variable, summer fixed), the best SCR can be achieved with the operation of VRFB with a maximum value of 95% but a lower value from 5-35 %. The lowest LCOE is achieved through a VRFB operating range from 5-95%, although LIB should be operated until the minimum SOC of 10 %, though the highest limit can be variable (60-90%). The lowest OBU is about 14 %, and the LIB range should stay at 40-80 %. Concerning Uc3 (winter fixed, summer variable), the lowest SOC could remain at 5 %, and the highest range is from 65-95 %. The OBU result is similar to the previously discussed case, and the lowest LCOE is achieved at a 10-90 % LIB range.

The increase of the lower limits and the decrease of higher limits of SOC range can benefit the battery's lifetime and simultaneously maintain or improve (not worsen) the KPIs (depending on the goal). This is the case of the 1-3 and 5 scenarios, where more than one SOC range offers a satisfactory (energy and economical) result and favours the operation far from the maximum boundaries.

In Scenario 5, VRFB single case presents the most similar energy KPIs to Scenario 1 and 2, although battery-related KPIs (with the indexing batteries names) are distinct. In terms of economic indicators, this scenario is not the best nor the worst. LIB single case presents itself in the middle of the energy indicators, except the case of FGU and EG, meaning the highest dependence from the grid. Through the observation of Table 4, this configuration presents the lowest investment, LCOE and SPB, and the highest NPV and IRR.

Generally, considering the methodology followed and defined inputs, the HESS system is only competitive when Scenario 3 is compared to Scenario 5-VRFB. Considering the current costs, none of the HESS configurations studied compete with Scenario 5-LIB. Despite this result, it must be considered that the lifetime of lithium batteries depends on the type of technology and its operation and use over the lifetime and that this work carried out its analysis for fifteen years. In case the lithium battery has to be replaced before the considered timeframe (another option for the replacement factor could be, for example, in the case of its energy capacity being less than 50% of the initial nominal energy capacity), the economic analysis must be updated, which will weigh on the indicators. This issue does not arise with the VRFB, whose lifetime is above the considered period.

Although the battery system cost decrease is expected, investing in HESS should be weighted for each application. Applications that find HESS investment competitive could rely on the determined energetic KPIs, which could compensate the investment (costs difference in using a single battery or a HESS). Those applications are related to enhancing the energy management strategy applied to the system, which depends on the goal of the application or service it aims to provide.

## 5. Conclusion

Current literature presents a lack of EMSs and power allocation that better concerns the economic and energy perspectives of HESS configurations, especially the LIB and VRFB for the building sector. The configuration LIB and VRFB (3.3 kW/ 9.8kWh and 5kW/ 60 kWh, respectively) set a favourable collaborative performance, given the complementary differences, allowing different combinations of power allocation and EMSs, answering to different problems, or improving case scenarios.

In this work, the EMSs of ESS and power allocation techniques of HESS are clarified with the proposal of five case scenarios, three of which are HESS approaches. The evaluation is done by calculating KPIs of single-ESS, and relevant analysis is addressed. Through the studied scenarios, one can conclude that the competitiveness of the ESS-single cases is possible with HESS, depending on the goal. It is possible to state that the VRFB-single case is improved by adding a LIB (HESS), Scenario 3. The seasonality factor, addressed in Scenario 4, could be a relevant approach in some geographical areas (e.g., Portugal), depending on the consumption and PV generation profiles and the exigency that the designed EMS makes of the battery use.

In general, the best LCOE results obtained (Table 4) show that these technologies still need improvement to reach grid parity. Additional cashflows of adding battery storage to a project are not quantifiable in the present analysis, as is the case of matters of security of supply frequency balancing issues, among others, considered crucial in the technology use and suitability.

When managing the HESS, the renewable energy sources' generation and consumption profiles significantly impact the project output and are highly dependent on the desired goal. The initial investment costs (CAPEX) of a battery system are relevant in the project analysis, and their costs are expected to be reduced in the following years due to the increasing adoption of VRE and massive application. The management of the overall electric system will include HESS management, sometimes with distinct EMSs competing for goals and different energy storage characteristics.

## 6. Future research work

The results obtained are based on a developed in-house tool that allows the overall microgrid assessment based on variation inputs. The analysis will improve results accuracy if the characterisation of the systems is improved, overall tool optimisation additions are achieved, and recent analysis costs are made. In this work, the representation of LIB lifetime is based on existing experience in literature, and a complete approach could be developed and tested in the future for further use.

Concerning Scenario 3, the LIB SOC measurement in real-time operation of LIB is needed. The investigation is on course to obtain more accurate methods of determining SOC. Without accurate SOC measurement, Scenario 3 results are affected. Despite the conditions of the Scenario 4 approach, each project should further assess the seasonality factor, depending on the geography and year. Other approaches to deal with seasons could be investigated, for example, for the consumption patterns within daily periods. Beyond the energy and economical approach followed in this work, the real-time implementation still needs to be addressed to complement the simulation and collect data using the microgrid of the University of Évora.

To better evaluate the benefits of investments in HESS, the simulated and other promising energy management controls should be achieved with the exploitation of each of these studied scenarios to reach the grid parity of HESS.

## Acknowledgements

The authors would like to thank Dr. Afonso Cavaco for reviewing this work. This work was co-funded through the COMPETE 2020 (Operational Program Competitiveness and Internationalisation) in the HyBRIDSTORAGE project with reference POCI-01-0247-FEDER-048270. It was also supported by INIESC – Infraestrutura Nacional de Investigação em Energia Solar de Concentração -, FCT / PO Alentejo / PO Lisboa, Candidatura: 22113 – INIESC AAC 01/SAICT/2016 (2017- 2021). The authors acknowledge the support of the ICT – Institute of Earth Sciences. This work was also supported by the PhD. scholarship (author Ana Foles) of FCT – Fundação para a Ciência e Tecnologia –, Portugal, with the reference SFRH/BD/147087/2019.